\documentclass[apj]{emulateapj}
\usepackage[colorlinks,linkcolor={blue},citecolor={blue},urlcolor={red}]{hyperref}
\usepackage{epsfig}
\usepackage{graphicx}
\usepackage{natbib}
\usepackage{amsmath}
\usepackage{amsfonts}
\usepackage{amssymb}

\newcommand{\myemail}{\email{sodi@kias.re.kr,leech@shao.ac.cn}}

\shorttitle{The environment of barred galaxies in the low-redshift Universe}
\shortauthors{Lin et al.}
 
\begin{document}

\title{The environment of barred galaxies in the low-redshift Universe}

\author{Ye Lin\altaffilmark{1}, Bernardo Cervantes Sodi\altaffilmark{1,2},
Cheng Li\altaffilmark{1}, Lixin Wang\altaffilmark{1},  and Enci Wang\altaffilmark{1}}
\myemail

\altaffiltext{1}{Partner   Group   of  the   Max   Planck  Institute   for
  Astrophysics  at  the  Shanghai  Astronomical  Observatory  and  Key
  Laboratory for Research in Galaxies and Cosmology of Chinese Academy
  of   Sciences,    Nandan   Road   80,    Shanghai   200030,   China}
\altaffiltext{2}{School of Physics, Korea Institute for Advanced Study, 
Dongdaemun-gu, Seoul 130-722, Republic of Korea}

\begin{abstract}

We present a study of the environment of barred galaxies using a volume-limited sample of over 30,000 galaxies drawn from the Sloan Digital Sky Survey. We use four different statistics to quantify the environment: the projected two-point cross-correlation function, the background-subtracted number count of neighbor galaxies, the overdensity of the local environment, and the membership of our galaxies to galaxy groups to segregate central and satellite systems. For barred galaxies as a whole, we find a very weak difference in all the quantities compared to unbarred galaxies of the control sample. When we split our sample into early- and late-type galaxies, we see a weak but significant trend for early-type galaxies with a bar to be more strongly clustered on scales from a few 100 kpc to 1 Mpc when compared to unbarred early-type galaxies. This indicates that the presence of a bar in early-type galaxies depends on the location within their host dark matter halos. This is confirmed by the group catalog in the sense that for early-types, the fraction of central galaxies is smaller if they have a bar. For late-type galaxies, we find fewer neighbors within $\sim$50 kpc around the barred galaxies when compared to unbarred galaxies form the control sample, suggesting that tidal forces from close companions suppress the formation/growth of bars. Finally, we find no obvious correlation between overdensity and the bars in our sample, showing that galactic bars are not obviously linked to the large-scale structure of the universe.

\end{abstract}

\section{Introduction}
\label{sec:introduction}

  The presence of bars in galaxies  is a feature that was noticed from
  the earliest studies of external galaxies
  \citep{Hubble-36,deVaucouleurs-63}. More recent  estimates  on the
  fraction  of  barred disk  galaxies  fluctuate  between  one- to  
  two-thirds, depending on  the method to identify bars  and the inclusion
  or             not              of             weak             bars
  \citep{Mulchaey-97,Knapen-Shlosman-Peletier-00,Eskridge-00,
    Marinova-Jogee-07,Sheth-08,Cameron-10, Lee-12a, Oh-Oh-Yi-12}. The
  large  abundance of  stellar bars  in local  galaxies  has triggered
  interest on  the involvement of these features  on secular evolution
  \citep{Sellwood-Wilkinson-93,  Kormendy-Kennicutt-04, Cheung-E-13}.   Given  that
  stellar  bars  break  the  radial  symmetry of  galaxies,  they  are
  particularly efficient in redistributing matter and angular momentum
  between  different components,  namely  stars, gas  and dark  matter
  \citep{Hohl-71,Sellwood-80,Tremaine-Weinberg-84,Weinberg-85,
    Debattista-Sellwood-00,Athanassoula-02,Athanassoula-03,Martinez-Valpuesta-Shlosman-Heller-06}.
  Clear examples of  redistribution of material driven by  bars is the
  fuel                  of                 gas                 inward
  \citep{Shlosman-Frank-Begelman-89,Friedli-Benz-93},   resulting   in
  accumulation  of  material that  might  be  used  as a  build-up  of
  disk-like             bulges             or            pseudo-bulges
  \citep{Kormendy-Kennicutt-04,Athanassoula-05,Heller-Shlosman-Athanassoula-07}. Given
  the collisional nature of  gas, this component can effectively lose
  energy        during       shocks       and        flow       inward
  \citep{Shlosman-Frank-Begelman-89,Friedli-Benz-93},  which  help  to
  explain  the high concentrations  of molecular  gas found  on barred
  galaxies  \citep{Sakamoto-99,Sheth-05},  as   well  as  the  younger
  stellar populations and higher star formation rates in the bulges of
  barred    galaxies     when    compared    with     unbarred    ones
  \citep[e.g.,][]{Coelho-Gadotti-11, Wang-12}. As a result, the overall
  structure  and morphology  of  barred galaxies  gets  shaped by  the
  effects of stellar bars.

At the same time, bar formation and evolution depends on many physical
properties of  hosting galaxies. In general, bars  are more frequently
found  in massive, red  galaxies with  prominent  bulges and  overall
early-type          morphologies          \citep{Sheth-08,Weinzirl-09,
  Hoyle-11,Masters-11,Lee-12a, Cheung-E-13}. Late-type, less massive, blue galaxies
also show the presence of stellar bars, but their frequency is low and
the     bars     they      exhibit     are     shorter     in     size
\citep{Elmegreen-Elmegreen-85,Erwin-05,Aguerri-Gonzalez-Garca-09,Lee-12a}. \citet{Masters-12}
found that  the bar fraction  is significantly lower in  gas-rich disk
galaxies than in gas-poor ones, a result in agreement with theoretical
expectations from \citet{Athanassoula-Machado-Rodionov-13}, where they
find in  their simulations that  bars form later in  gas-rich systems,
and after they  form, they grow more slowly  than in poor-gas systems,
hence predicting that the bar  likelihood should be higher in galaxies
with low  gas content.  Using  numerical simulations of  spinning dark
matter  halos,  \citet{Long-Shlosman-Heller-14}  demonstrate that  bar
growth in  strength and size,  is strongly suppressed on  systems with
spin  parameter   $\lambda  \gtrsim  0.03$,  giving   support  to  the
observational finding by \citet{Cervantes-Sodi-13}, where they reached
this same  conclusion estimating $\lambda$ for the  same galaxy sample
employed  on the  present  study,  and finding  that  strong bars  are
confined  to reside  in galaxies  with low  to intermediate  values of
$\lambda$,  while  weak  bars  are  preferentially  found  in  rapidly
spinning systems. More recently, \citet{Cervantes-Sodi-14} explored
the dependence of the bar fraction on the stellar-to-halo mass
ratio (M$_{\mathrm{*}}/$M$_{\mathrm{h}}$) for central disk galaxies finding,
in agreement with theoretical studies \citep{Ostriker-Peebles-73,
Efstathiou-82, Christodoulou-95, DeBuhr-12}, that bars are more common
in galaxies with high M$_{\mathrm{*}}/$M$_{\mathrm{h}}$ values, and
exploring the bar fraction in the M$_{\mathrm{h}}$ vs M$_{\mathrm{*}}$
plane, they find that the dependence is stronger considering a relation
with the form $f_{\mathrm{bar}}=f_{\mathrm{bar}}$(M$_{\mathrm{*}}^{\alpha}/$M$_{\mathrm{h}}$)
with $\alpha=1.5$.

Although secular  evolution has  proven to be  an effective  driver of
evolution,  it is  by no  means the  only way  in which  a  galaxy can
transform.   Regarding  the  formation   and  growth  of  bars,  early
simulations       of      interacting      disk       galaxies      by
\citet{Noguchi-87,Noguchi-88}  show that  self-gravitating  disks that
were  expected to be  stable against  the development  of a  bar, once
perturbed by the tidal force  of a companion, develop prominent spiral
structures, that at  the center resemble bar structures  that are able
to induce gas  infall into the nucleus when  gas dissipates energy via
cloud-cloud   collisions.     Similar   results   are    reported   by
\citet{Byrd-Valtonen-90}  for the case  of tidal  forces exerted  by a
cluster   potential  on   simulated  disk   galaxies.    Although  the
configuration of  the interaction changes the final  properties of the
bar, some  groups found that in  general, it enhances  the formation of
the  bar  \citep{Gerin-Combes-Athanassoula-90,Miwa-Noguchi-98},  while
others report  that the  opposite is also  possible: the bar  can loose
angular  momentum  and  mass  and  debilitate  after  the  interaction
\citep{Sundin-Sundelius-91,Sundin-Donner-Sundelius-93}.  There seem to
be  also  controversy about  the  different  effects  of prograde  and
retrograde                                                 interactions
\citep{Romano-Daz-08,Aguerri-Gonzalez-Garca-09}.

Ultimately, the  effect of environment on the  formation and evolution
of  bars  needs  to   be  addressed  through  observational  evidence.
\citet{Thompson-81} found a larger  fraction of barred galaxies in the
core of  the Coma cluster  and \citet{Giuricin-93} reported  a higher
recurrence  of  early-type  barred   spirals  in  high  local  density
environments.  Later  studies of local \citep{Andersen-96,Eskridge-00}
and intermediate-redshift clusters \citep{Barazza-09}  showed similar
findings and, more recently, \citet{Mendez-Abreu-12} proved that
by taking into account the different luminosity ranges, it is possible to
detect different effects of the environment for luminous and faint galaxies
located in clusters or in the field. They propose that interactions in
bright disk galaxies that are stable enough against interactions trigger
bar formation, while for faint galaxies, interactions are strong enough to
heat the disk inhibiting bar formation. 

The opposite result--no evidence  for a dependence of bar frequency
on environment,  comparing field and cluster galaxies--is also widely
reported \citep{vandenBergh-02,
  Marinova-09,Aguerri-Gonzalez-Garca-09,Mendez-Abreu-Sanchez-Janssen-Aguerri-10,
  Giordano-11,Marinova-12}. When the samples include only S0 galaxies,
most of the  time, a secondary dependence on  environment is reported,
with   a  higher  frequency   of  barred   S0  galaxies   in  clusters
\citep{Barway-Wadadekar-Kembhavi-11}  and  an   increase  in  the  bar
fraction toward the cluster core \citep{Lansbury-Lucey-Smith-14}.

By  computing  projected  redshift-space two-point  cross-correlation
functions (2PCCF) for a sample of nearly 1,000 galaxies from the SDSS,
\citet[][hereafter  L09]{Li-09} studied  the clustering  properties of
barred  galaxies finding that  the clustering  of barred  and unbarred
galaxies of similar stellar  mass is indistinguishable and no evidence
that  bar formation  is  promoted by  mergers  or interactions.   More
recently, \citet{Skibba-12}, also using clustering statistics, reported
a positive  correlation for bulge-dominated and barred  galaxies to be
found in denser  environments on scales of 150 kpc to  a few megaparsecs than
their unbarred and disk-dominated  counterparts and argued that barred
galaxies are often central galaxies in low-mass dark matter halos or
satellites in more massive ones.

\citet{Lee-12a} also investigated the  dependence of the barred galaxy
fraction on environment,  finding that once mass and  color are fixed,
the  bar fraction  is independent  of the  background density,  but an
influence of the nearest neighbor  appears when the separation to the
nearest  neighbor is less  than 0.1  times the  virial radius  of the
neighbor.   As  the  distance  decreases,  the  bar  fraction  drops,
regardless of  the morphology of the neighbor--a possible indication
that  strong   tidal  interactions  destroy  bars   or  prevent  their
growth. Similar results  were presented by \citet{Mendez-Hernandez-11}
comparing reduced samples of  isolated galaxies and galaxies in pairs,
showing  that the  subsample  of  galaxies in  pairs  presented a  bar
fraction of only 20\%, while that for isolated systems was 43\%. Using
a much  larger galaxy  sample, \citet{Casteels-13} reported  a similar
result  using  the  Galaxy Zoo  2  sample,  with  a decrease  for  the
likelihood of  identifying a bar  in pairs with separations  $\leq$30
$h^{-1}$ kpc.

In this  paper, we study the  environment of galaxies with  bars in the
local universe, extending the analysis  of L09 by using the new sample
provided by \citet{Lee-12a}, which is  30 times larger than the sample
used by L09. When compared to  previous studies of the same topic, our
approach differs in the following ways.
\begin{itemize}
  \item Instead of computing the auto-correlation of the barred galaxy
    sample,  we  compute  the  cross-correlation of  our  sample  with
    respect to reference samples of the general galaxy population from
    scales of a  few tens of kiloparsecs up  to a few tens of  megaparsecs. This takes
    advantage  of  the much  larger  sample  size  and volume  of  the
    reference  samples,  allowing  us  to substantially  increase  the
    signal-to-noise  ratio in  our results,  as  well as  allowing us to study
    the scale dependence of the clustering of barred galaxies in detail.
   \item  We wish  to isolate  the  link between  environment and  the
     presence  or absence  of a  bar in  galaxies, so  the environment
     measurements of  the barred galaxies are always  compared to those
     of control samples of  unbarred galaxies that are closely matched
     in  galaxy  properties known to  be  correlated  with
     environment.  According to previous  studies of the clustering of
     galaxies   \citep[e.g.][]{Li-06a},  we  consider   stellar  mass,
     optical color  and internal structure, quantified  by the surface
     stellar mass  density and the concentration of  the stellar light
     distribution, when constructing the control samples.
 \item  In  addition to  computing  the  2PCCFs,  we use  three  other
   statistics to characterize the  environment of our galaxies.  These
   include  (1)  the   background-subtracted  average  neighbor  counts
   ($N_c$) around  the barred and  unbarred galaxies as a  function of
   the projected distance to  neighboring galaxies, (2) the overdensity
   ($\delta$) of  the local environment  of our galaxies  estimated on
   $\sim3$ Mpc scale, and (3) the membership of our galaxies in the SDSS
   group  systems   identified  by  \citet{Yang-07}.    The  different
   statistics  have   their  pros  and  cons   in  quantifying  galaxy
   environment,  and provide  measurements that  are  complementary to
   each other.
\end{itemize}

This  paper  is  organized  as  follows.   Section  2  gives  a  brief
description of the  volume-limited sample, as well as  the control and
reference samples used  in this study.  In Section  3, we describe the
methods used  to characterize the  environment of the galaxies  in our
sample.  In  Section 4, we  present our general results.   Finally, in
Section  5  we summarize  our  results  and  present our  conclusions.
Throughout this  paper, we  assume a cosmology  with a  density parameter of
$\Omega  =  0.3$,  cosmological  constant of $\Omega_\Lambda  =0.7$,  and
Hubble constant of $H_0$ = 70 km s$^{-1}$ Mpc$^{-1}$.

\section{Data}
\subsection{The sample of barred galaxies}
\label{sec:data}

The sample of galaxies with bar identification used in our study comes
from  a  previous  work  by  \citet{Lee-12a}. Here  we  give  a  brief
description of the sample selection and morphology classification, and
the reader is referred to  \citet{Lee-12a} for the full description of
the   sample    and   comparisons   with    previous   classifications
\citep{deVaucouleurs-91,Nair-Abraham-10} and Park \& Lee (2014) for the
publicly available catalog.     The    sample    is    a
volume-limited sample  selected from the Korea  Institute for Advanced
Study  Value-Added Galaxy  Catalog, which  was constructed  from the
SDSS DR7 \citep{Abazajian-09} by \citet{Choi-Han-Kim-10} and consists
of  33,391 galaxies  with  $r$-band absolute  magnitudes brighter  than
$M_{r} =$  --19.5 + 5log$h$ and  spectroscopically measured redshift in
the range 0.02 $\leq z \leq$ 0.05489.  The segregation into early- and
late-type   galaxies   is done by  adopting   the   prescription   by
\citet{Park-Choi-05},  where galaxies are  divided according  to their
morphology  in color  versus  color gradient  and concentration  index
space  plus an  additional visual  inspection.  The  identification of
bars is  done by  visual inspection of  $g+r+i$ combined  color images
from  SDSS. Early-type  galaxies are  classified as  either  barred or
unbarred galaxies. For the case  of late-type galaxies, given that the
classification is more robust for face-on systems, we limit our sample
to  galaxies  with  a  minor-to-major  axis  ratio of  $b/a>0.6$.   For
late-type galaxies presenting a bar,  the bar is further classified as
{\em strong} or {\em weak} according to its size--if the bar is larger
than one quarter the size of the hosting galaxy, it is classified as a
strong bar, otherwise it becomes a weak bar.
Given that the bar fraction estimated by visual inspection of
optical images is statistically the same as that estimated using
automated methods such as ellipse fitting \citep{Menendez-Delmestre-07,Sheth-08}
and that the computation of the bar fraction gives the same result using
near-infrared images as it does when using optical \citep{Whyte-02,Menendez-Delmestre-07,
Sheth-08}, we are confident that our results are reproducible even if
the bar detection is obtained by using a different method.

We  extract the  physical quantities  required for  our work  from two
additional  catalogs:  the New  York  University  Value Added  Galactic
Catalog \citep[NYU-VAGC;][]{Blanton-05}
\footnote{http://sdss.physics.nyu.edu/vagc/}  and   the  MPA/JHU  SDSS
database \citep{Kauffmann-03, Brinchmann-04}
\footnote{http://www.mpa-garching.mpg.de/SDSS/},  both  of  which  are
based on SDSS DR7 and  publicly available.  These include stellar mass
$M_\ast$, stellar  surface mass density $\mu_\ast$,  color $(g-r)$ and
concentration index  $C$. The stellar  mass of a galaxy  was estimated
based on its redshift and  the five-band SDSS photometry following the
methodology detailed in \citet{Kauffmann-03}
\footnote{Note  that  the  masses  included  in  the  current  MPA/JHU
  database are  based on fits to  photometry and so are  not identical to
  those  of \citet{Kauffmann-03}  where  the masses  are obtained  from
  spectroscopic                      indices.                      See
  http://www.mpa-garching.mpg.de/SDSS/DR7/mass\_comp.html for detailed
  comparisons  between the  two  masses.}.  The  surface stellar  mass
density of a galaxy is defined by $\mu_\ast=0.5M_\ast/\pi R_{50,z}^2$,
where $R_{50,z}$ is the radius enclosing 50\% of the $z$-band flux, and
the  concentration index  $C=R_{90,r}/R_{50,r}$, where  $R_{90,r}$ and
$R_{50,r}$ are the radii enclosing 90\% and 50\% of the total light in
the $r$-band  image of  the galaxy. Our  final sample  includes 17,839
unbarred  and 3,749  barred  galaxies. In  order  to distinguish  this
sample  from subsequent control  samples, we  will refer  to it  as C0
sample,  which we  will divide  into eight  different subsamples:  AU (AB)
includes  all  unbarred  (barred)  galaxies,  EU  (EB)  includes  only
unbarred  (barred) early-type galaxies,  LU (LB)  includes only
unbarred (barred) late-type galaxies, SB includes only strongly barred
late-type galaxies and WB includes weakly barred late-type galaxies.

\subsection{Control samples of unbarred galaxies}

Previous studies have shown  that barred and unbarred galaxies present
different   stellar    mass,   color,   and    surface   mass   density
distributions,and at the same time the clustering of galaxies strongly
depends  on all  these properties  \citep[e.g.][]{Li-06a}.In  order to
isolate the  effect of environment on  bars, we construct  a series of
control samples of  unbarred galaxies, and in what  follows we compare
our results of  each barred-galaxy subsample always with  those of the
corresponding  control  sample.   We  compile  two  different  control
samples:  C2 encompasses galaxies  with matched  stellar mass  and color
between  barred  and  unbarred  galaxies  and C3  matches  an  extra
quantity--the  stellar surface mass  density.  The  matching tolerances
are $\triangle$log$M_* \leq$ 0.08 , $\triangle(g-r)\leq$ 0.025,
$\triangle$log$\mu_*\leq$ 0.08.

\begin{figure*}
  \begin{center}
    \epsfig{figure=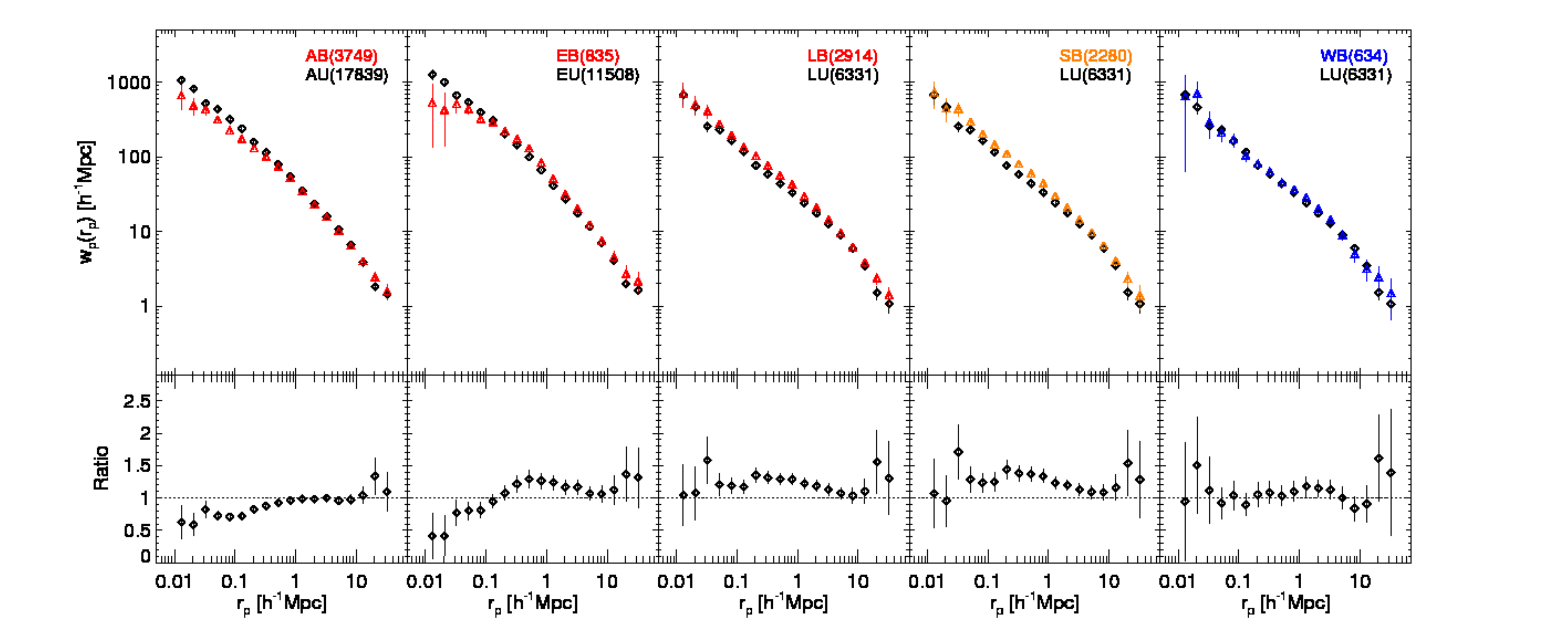,width=\hsize}
    \caption{Projected  2PCCF for  barred  $w_p^{bar} (r_p)$  and
      unbarred  $w_p^{unbar}  (r_p)$  galaxies (top  panels)  and
      $w_p^{bar} (r_p)$ to $w_p^{unbar} (r_p)$ ratio (bottom
      panels) for the different subsamples of our parent galaxy sample
      C0. Panels  from left  to right correspond  to the  whole galaxy
      sample of  barred (AB orange triangles) plus  unbarred (AU black
      diamonds) galaxies, early-types barred (EB orange triangles) and
      unbarred  (EU  black  diamonds),  late-types barred  (LB  orange
      triangles)  and unbarred  (LU black  diamonds),  strongly barred
      late-types  (SB  yellow triangles)  and  LU,  and weakly  barred
      late-types (WB blue triangles) and LU.}
  \label{fig:2PCCF_C0}
  \end{center}
\end{figure*}

\subsection{Reference galaxy samples}

We  have  constructed two  reference  samples  from  NYU-VAGC: a  {\em
  spectroscopic}  reference  sample,  which  is used  to  compute  the
projected two-point  cross-correlation function(2PCCF) $w_p(r_p)$, and
a  {\em photometric}  reference  sample, which  is  used to  calculate
counts  of close neighbors  around our  barred and  unbarred galaxies.
The spectroscopic  reference sample  is constructed from  version {\tt
  dr72} of  the NYU-VAGC,  the current version  built on the  SDSS DR7
data,  by  selecting all  galaxies  with  $r$-band apparent  magnitude
corrected  for foreground extinction  $r<17.6$ within  the restricted
redshift range  of $0.01\leq  z \leq 0.2$  and the  absolute magnitude
range $-23\leq M_{0.1_r} \leq  -17$.  Here $M_{0.1_r}$ is the $r$-band
absolute magnitude  corrected to its value  at $z =  0.1$.  With these
selection criteria, we are left  with about half a million galaxies in
our spectroscopic reference  sample.  The photometric reference sample
is  also constructed from  NYU-VAGC version  {\tt dr72}  by selecting
galaxies with  $r$-band apparent magnitude down to  $r<21$. The sample
includes about 2.5 million galaxies.

\section{Methodology}

In  this  section we  briefly  describe the  methods  we  will use  to
characterize the  environment of the  galaxies in our samples.   For a
full description of  these methods, we refer the  reader to the source
papers      where      they      are     presented      in      detail
\citep{Li-06a,Li-06b,Yang-07,Jasche-10}.

\subsection{Projected two-point cross-correlation function}

We  use  the   projected  redshift-space 
2PCCF, $w_p(r_p)$, to quantify the clustering properties of
our galaxies.  For  this we have constructed a  random sample that has
the  same  selection effects  as  the  spectroscopic reference  sample
following the  method described in  \citet{Li-06a}. We cross-correlate
each of our  samples of barred galaxies, as  well as the corresponding
control  sample  of  unbarred  galaxies,  with  respect  to  both  the
spectroscopic reference sample and  the random sample, and then define
$w_p(r_p)$  as a  function of  the projected  separation $r_p$  by the
ratio of the  two pair counts minus one. A  careful correction for the
effect  of fiber collisions  is implemented  by comparing  the angular
2PCF of the  spectroscopic sample with that of  the parent photometric
sample  \citep[see][for details]{Li-06b}.   Errors  of the  $w_p(r_p)$
measurements  are  estimated   using  the bootstrap  resampling  technique
\citep{Barrow-Bhavsar-Sonoda-84}, for  which a total  of 100 bootstrap
samples  are generated  based on  the real  sample. Details  about our
methodology   for  computing   the  correlation   functions   and  for
constructing the random sample can be found in \citet{Li-06a}.

\subsection{Close neighbor counts}

We count the number of galaxies in the photometric reference sample in
the vicinity of the galaxies in our samples, and we make a statistical
correction  for the effect  of chance  projections by  subtracting the
average count  around randomly placed galaxies.  When  compared to the
2PCCF,  close neighbor counts  are not  affected by  incompleteness on
small scales  caused by the  fiber collisions. In addition,  since the
photometric sample  is much deeper  than the spectroscopic  sample, by
computing  background-subtracted neighbor  counts  in the  photometric
sample one  is able to  include much fainter companion  galaxies, thus
probing the  effect of close companions  over a broader  range of mass
ratios.

\begin{figure*}[t]
  \begin{center}
    \epsfig{figure=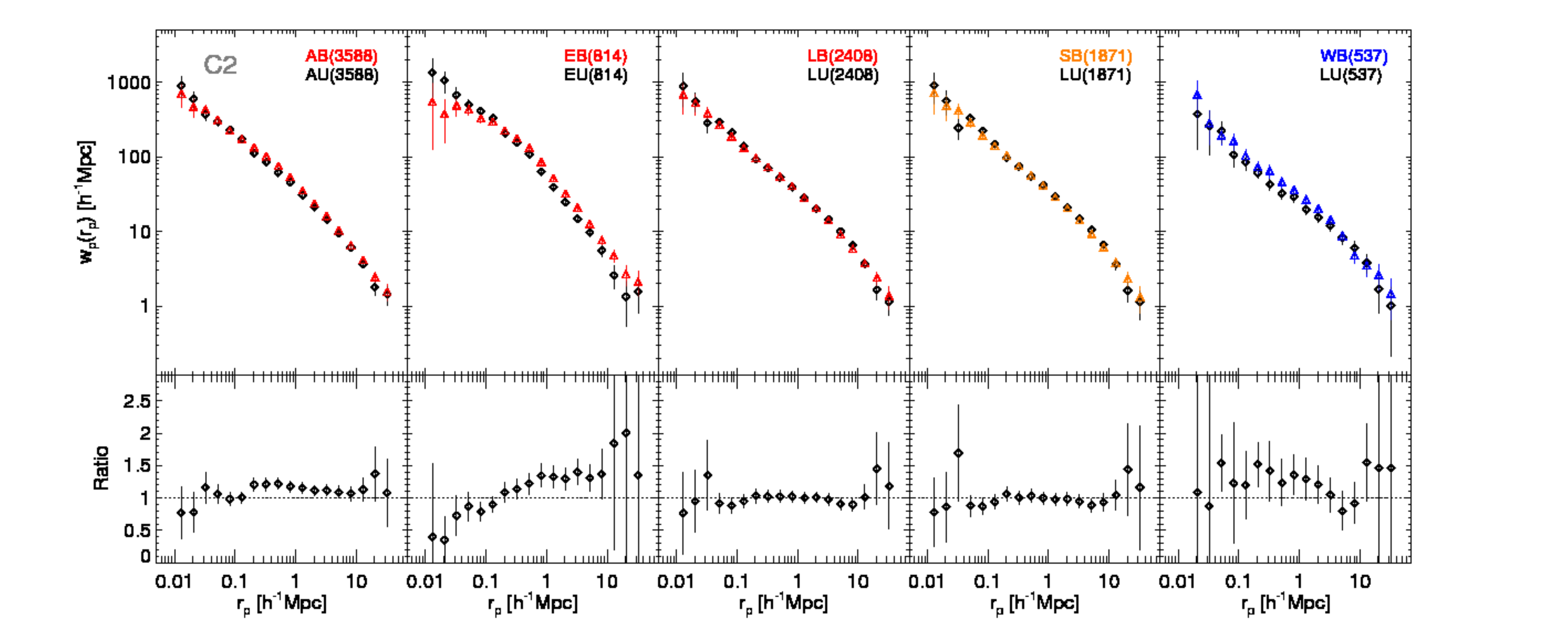,width=\hsize}
  \end{center}
    \caption{Same as  the Figure 1,  but using control sample  C2 where
      the galaxies  share a common distribution of  stellar mass $M_*$
      and color g$-$r.  }
\label{fig:2PCCF_C2}
\end{figure*}

\subsection{Group catalog}

To account for  the environment of galaxies in groups,  we make use of
the group catalog of \citet{Yang-07}, constructed from the sample {\tt
  dr72} of the NYU-VAGC.  The  groups of galaxies are identified using
a modified version of the halo-based group-finding algorithm developed
in \citet{Yang-05} and applied to a sample with redshifts in the range
0.01 $\leq z  \leq$ 0.20 and with a  redshift completeness greater than
0.7. From this  extensive catalog, we use only  galaxies identified in
groups with  at least  three members,  from which we  are left  with a
final sample of  nearly 2,000 galaxies. For each  group system, we take
the  most  massive galaxy  as  the central  galaxy,  and  the rest  as
satellites.

\subsection{Large-scale overdensity}

To account  for the  environment at large  scales ,  we make use  of a
non-linear, non-Gaussian, full  Bayesian large-scale structure analysis
of   the   cosmic  density   field   based   on   the  SDSS   DR7   by
\citet{Jasche-10}, where the authors use a Bayesian sampling algorithm
in  order  to  obtain  the extremely  high  dimensional  non-Gaussian,
non-linear  log-normal Poissonian  posterior  of the  three-dimensional
density field.  The reconstruction of the field is made over a 750 Mpc
cube, taking  into account the angular and  radial selection functions
of  the  SDSS, and  a  proper treatment  of  a  Poissonian shot  noise
contribution for an effective resolution of the order of $\sim$3 Mpc.

\section{Results}

\subsection{Cross-correlation functions}

\begin{figure*}[t]
  \begin{center}
    \epsfig{figure=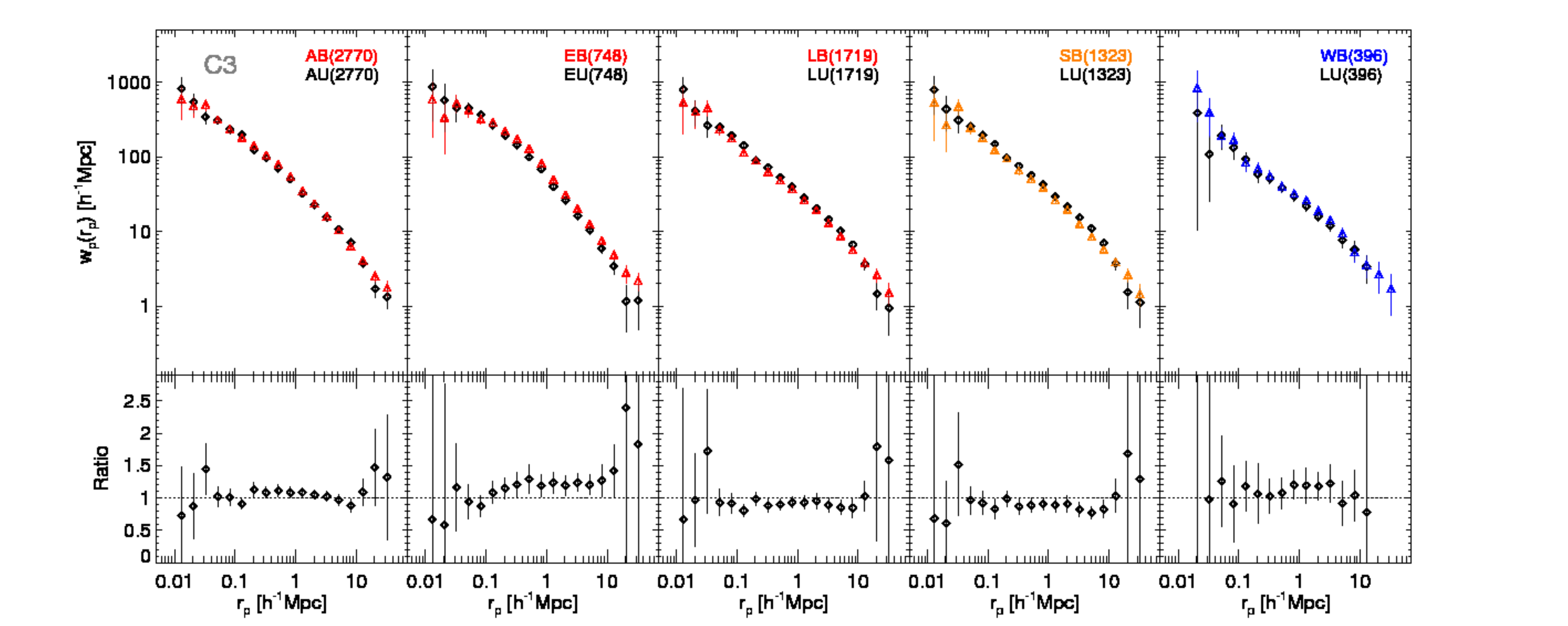,width=\hsize}
  \end{center}
    \caption{ Same as  the Figure 1, but using  control sample C3 where
      the galaxies share a  common distribution of stellar mass $M_*$,
      color g$-$r, and stellar surface mass density $\mu_*$.  }
\label{fig:2PCCF_C3}
\end{figure*}

\begin{figure*}[t]
  \begin{center}
    \epsfig{figure=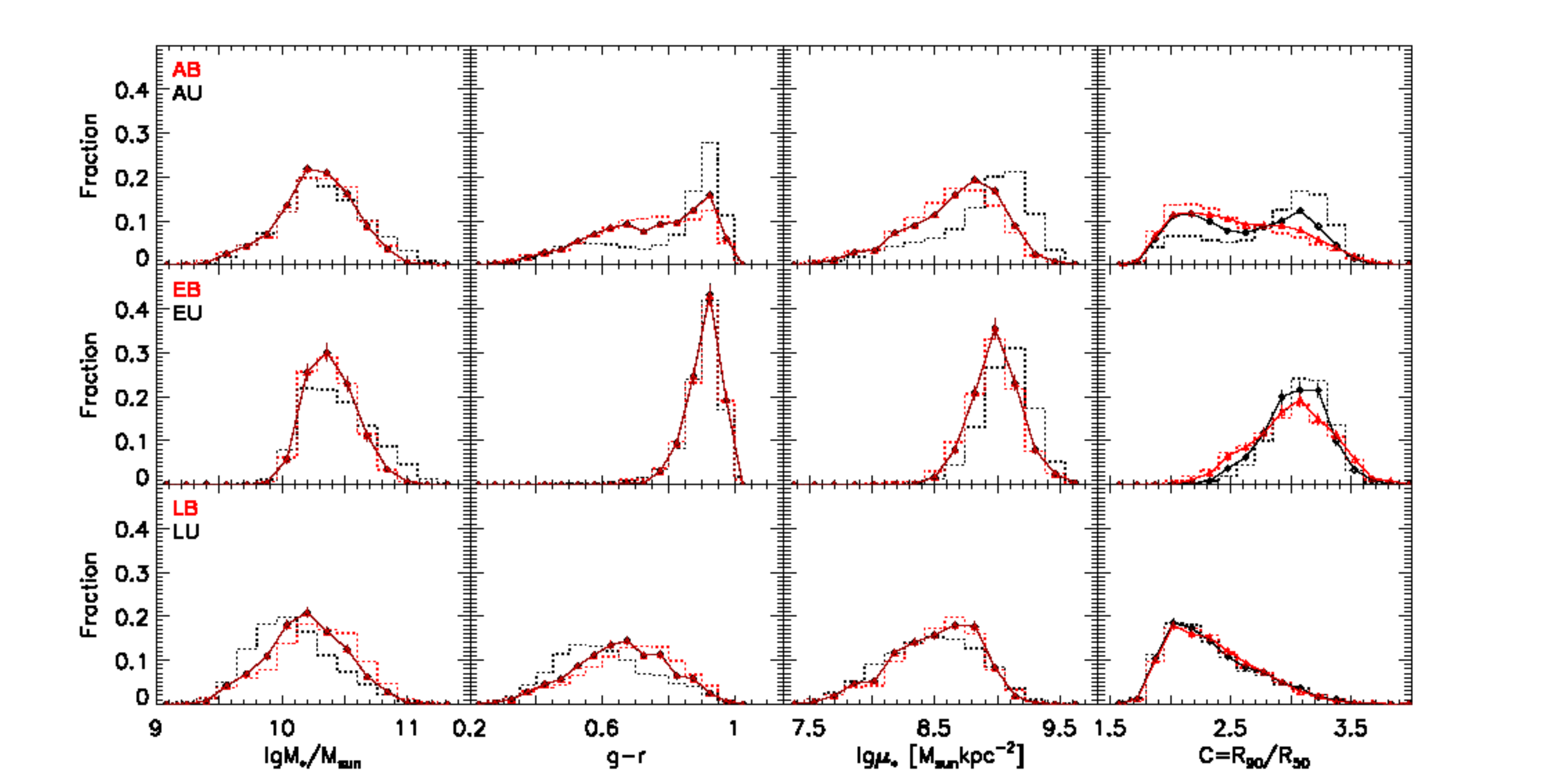,width=\hsize}
  \end{center}
    \caption{Histograms for  stellar mass $M_*$,  $g-r$ color, stellar
      surface mass  density $\mu_*$ and the concentration  index C for
      control samples  C0 (dotted lines)  and C3 (solid  lines). The first
      line corresponds to the whole sample, the second line to early-types,
      and the third line to late-types.}
\label{fig:sample_properties}
\end{figure*}

\begin{figure}[t]
  \begin{center}
    \epsfig{figure=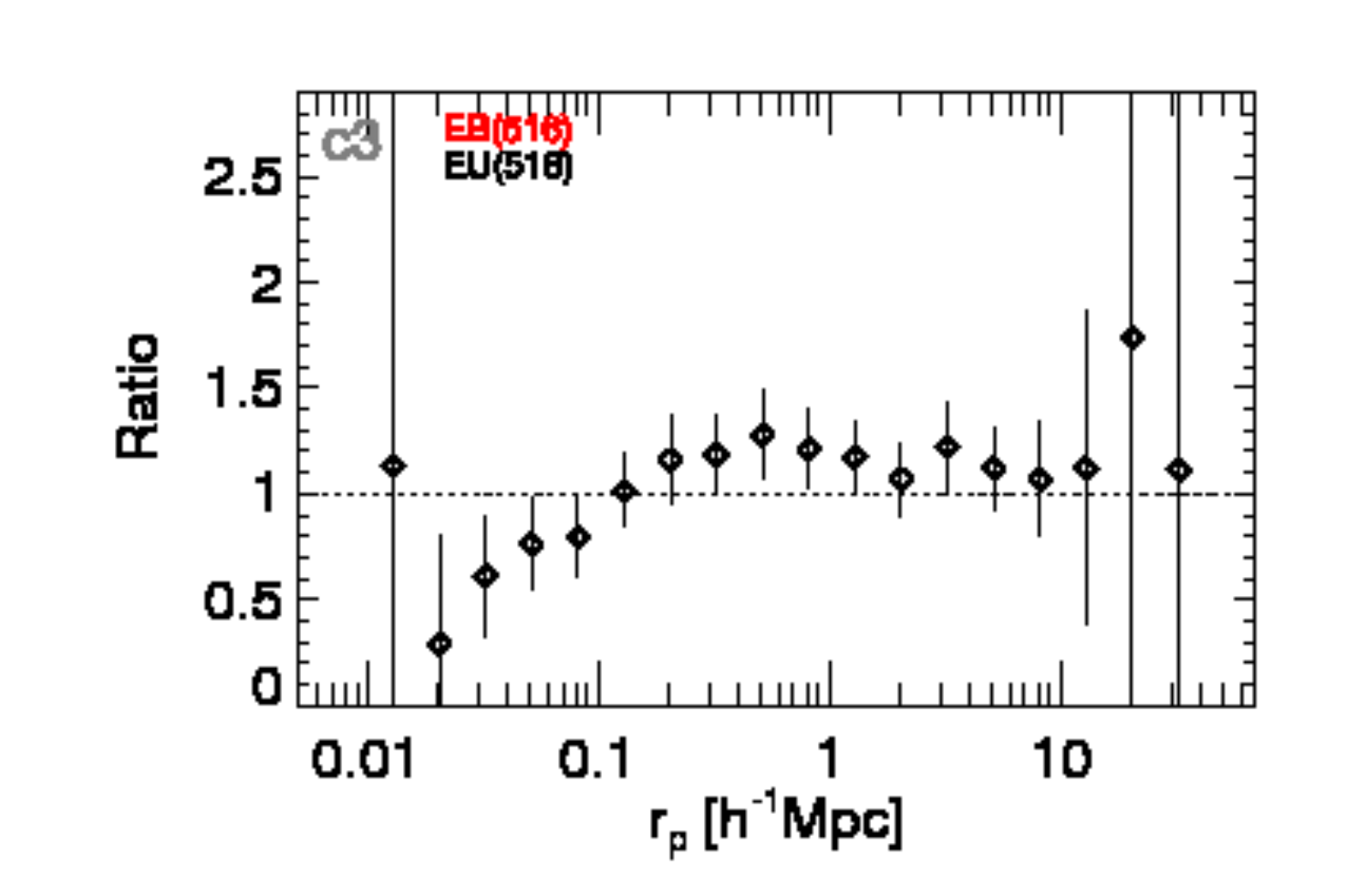,width=\hsize}
  \end{center}
    \caption{Same  as the  Figure 1,  but using  control sample  C3 for
      early-type galaxies with 2.8$\leq C \leq$ 3.5.  }
\label{fig:2PCCF_con}
\end{figure}
  
  \begin{figure*}[]
  \begin{center}
    \epsfig{figure=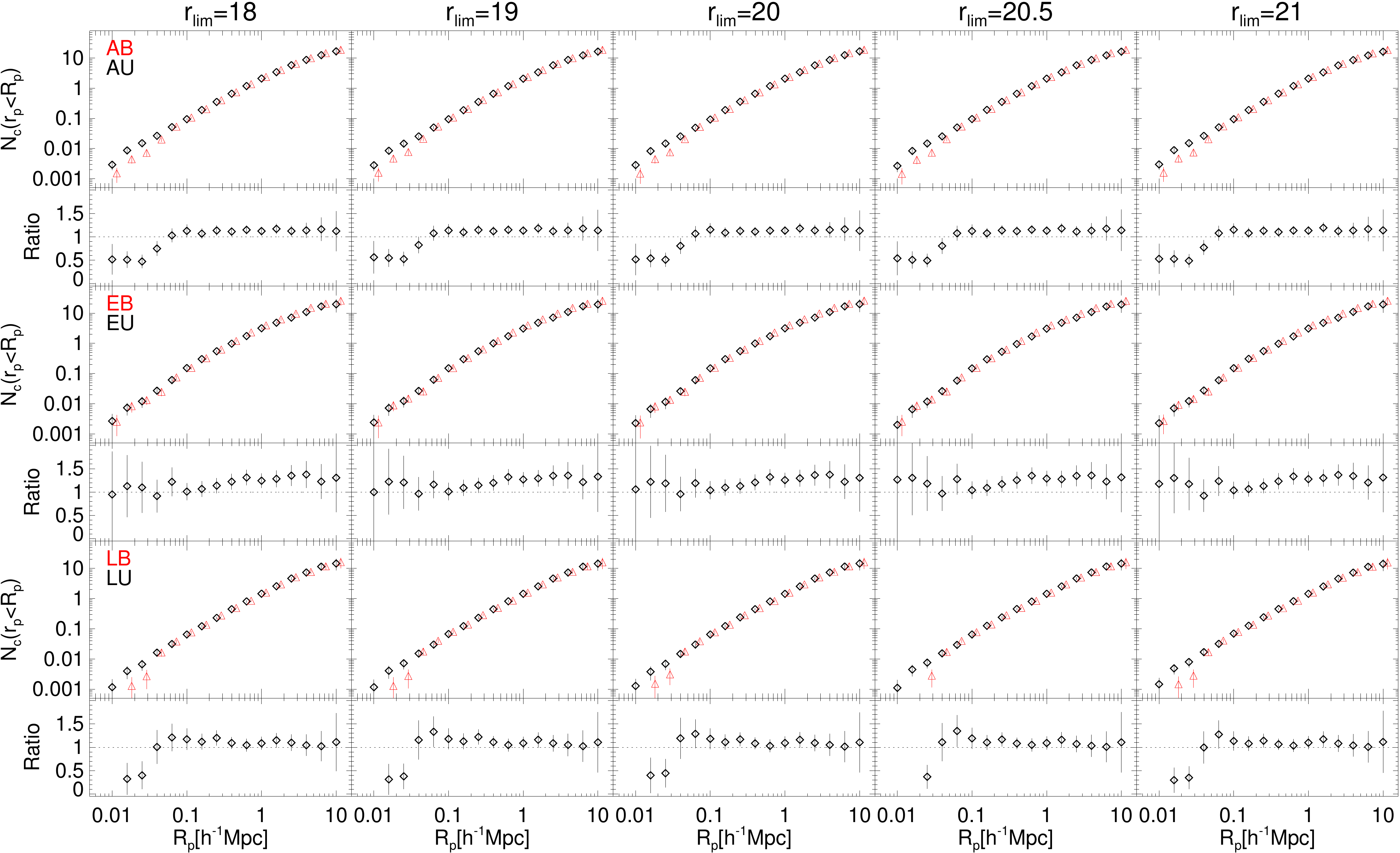,width=\textwidth}
  \end{center}
    \caption{Average  counts  of galaxies  in  the photometric  sample
      within a given  projected radius R$_P$ from the  galaxies in the
      C3  subsamples.   Each line  corresponds  to different  apparent
      magnitude  limits in the $r$ band  ($r\leq$ 18,19,20,20.5,21)  for the
      galaxies in the photometric sample.}
  \label{fig:nc}
\end{figure*}

In  Figure~\ref{fig:2PCCF_C0}, we show  the 2PCCF  $w_p(r_p)$ measured
for our main  galaxy sample C0. In each panel  we present $w_p(r)$ for
barred $w_p(r)^{bar}$ and  unbarred $w_p(r)^{unbar}$ galaxies for the
different  subsamples in consideration  (from left  to right:  all the
galaxies  in  the  sample,  early-types, late-types,  strongly  barred
late-types and weakly barred late-types).  For clarity we also include
the amplitude ratio $w_p(r)^{bar}$ to $w_p(r)^{unbar}$ for each case.

At  first sight,  it is  clear that  the clustering  of our  barred and
unbarred  samples are very  similar for  all the  subsamples, although
some differences appear when  looking carefully at each case. Starting
with  the extreme left  panel, which  includes all  the galaxies  in our
sample, we  notice that unbarred galaxies are  more strongly clustered
than barred  ones on  scales below $\sim$1  Mpc.  If we  include only
early-types  in  our analysis  (second  panel),  the clustering  ratio
between  barred and  unbarred systems  changes, presenting  a  pick at
around 1 Mpc and positive values  at larger scales. Only on scales
bellow 100 kpc does the ratio becomes less than unity. When we turn to
the case of late-type  galaxies (middle panel), barred galaxies appear
to  be more strongly  clustered than  their unbarred  counterparts at
almost all the scales probed.  If we distinguish between strong (fourth
panel) and  weak (fifth panel) bars,  we see that the  signal seen for
the case  of late-type  galaxies mostly comes  from the  clustering of
strongly barred  galaxies, which is natural given that they are more 
numerous in our sample than weakly barred galaxies.

At  this point,  Figure~\ref{fig:2PCCF_C0} seems  to indicate  that the
clustering of barred  and unbarred galaxies is different,  but we need
to be careful  to drive conclusions at this  stage given the following
two  facts.  First, previous  studies of  galaxy clustering  have well
established  that  clustering strongly  depends  on galaxy  properties
including  mass, color  and morphological  type \citep[e.g.][]{Li-06a,
  Zehavi-11}.   Second, it  is also  well known  that the  presence or
absence of bars  is not the only morphological  difference between our
barred   and   unbarred  samples.    For   instance,  previous   works
\citep{Barazza-Jogee-Marinova-08,Nair-Abraham-10,Lee-12a}   show  that
the fraction  of barred  galaxies is higher  in massive  red galaxies
than  in less-massive blue  systems.  Massive  red galaxies  are more
strongly clustered than less-massive, blue galaxies.  To normalize out
this  effect  we make  use  of  our control  samples  C2  and C3.   As
described  in Section  2.1,  for  the case  of  C2 we  have
restricted our samples  so that each subsample of  barred galaxies has
the same distribution  in stellar mass and color  as the corresponding
subsample of unbarred  galaxies, while for the case  C3 the subsamples
of  barred   and  unbarred   galaxies  being  compared   are additionally  required
to have the  same distribution  in stellar  surface mass
density.

Measurements  of the  2PCCF for  our C2  and C3  samples are  shown in
Figures ~\ref{fig:2PCCF_C2}  and~\ref{fig:2PCCF_C3}, respectively. For
late-type galaxies, the  different clustering behaviors between barred
and unbarred galaxies seen in Figure~\ref{fig:2PCCF_C0} disappear when
the samples being compared are  closely matched in stellar mass, color
and  surface   mass  density,  with  the   ratio  $w_p(r_p)^{bar}$  to
$w_p(r_p)^{unbar}$  consistent with  unity  within error  bars on  all
scales probed. The same result holds if we further split our sample of
barred  late-type  galaxies  into  strongly barred  and  weakly barred
galaxies according to the size of their bars.  This is consistent with
the expectation  that the  relatively strong clustering  of late-type,
barred galaxies  seen for the full  sample (i.e. sample  C0) is indeed
attributed  to  the fact  that  the  subset  of barred  galaxies  with
late-type  morphology in  C0 are  more massive  and redder,  thus more
strongly clustered, when compared  to the unbarred late-type galaxies.
This result implies  that the presence of a  bar in late-type galaxies
is not obviously linked with their clustering properties as quantified
by the 2PCCF, confirming the earlier finding of \citet{Li-09} that was
based on a much smaller sample.

\begin{figure*}[t]
  \begin{center}
    \epsfig{figure=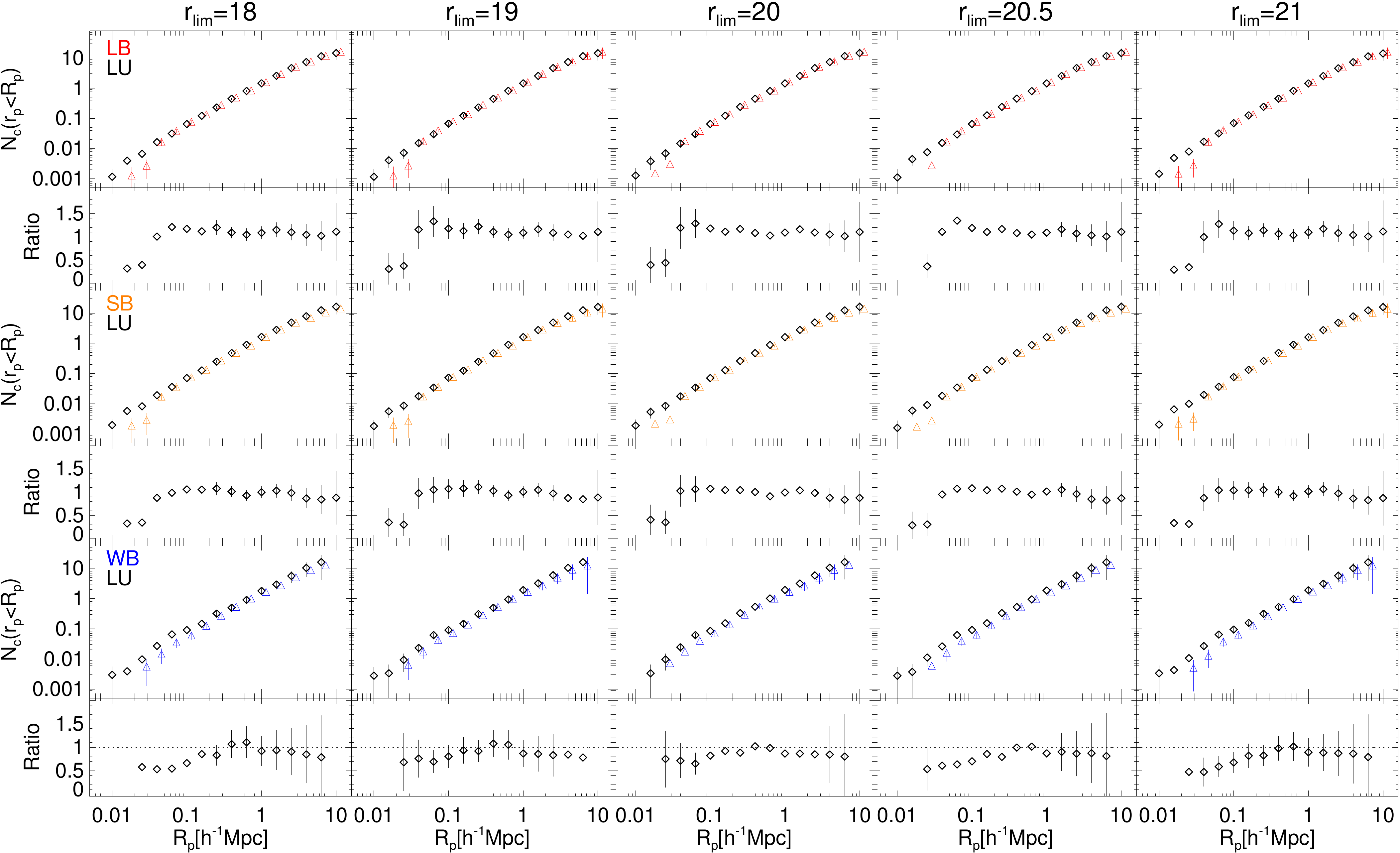,width=\textwidth}
  \end{center}
    \caption{Ratio of  average neighbor  counts of barred  to unbarred
      galaxies  for C3  within  a given  apparent  magnitude limit  in the
      $r$ band ($r\leq$ 18,19,20,20.5,21).  }
  \label{fig:nc_ratio}
\end{figure*}

For  early-type galaxies,  in contrast,  the difference  in  the 2PCCF
between barred  and unbarred galaxies is still  significantly seen for
scales larger  than a few hundred kiloparsecs, even when the  samples are closely
matched in mass, color, and  surface mass density. Considering that the
clustering of galaxies depends not exclusively on these parameters, we
compare   the distributions  of
stellar  mass, g$-$r  color, surface  mass density,  and concentration
index $C$ for samples C0 and  C3  in  Figure~\ref{fig:sample_properties}. In the case of C3, the distributions
for $M_*$, g$-$r and  $\mu_*$ are indistinguishable between barred and
unbarred  galaxies, regardless  of their  morphological type,  but the
concentration  index $C$  follows different  statistical distributions
between  barred  and unbarred  systems,  especially  for  the case  or
early-type galaxies (see the  rightmost panels).  In order to further
isolate  the  link between  bars  and  the  clustering of  early-type
galaxies,  we remove  the  dependence of  clustering on  concentration
index from our analysis by  restricting ourselves to a narrow range of
concentration  index 2.8$\leq  C \leq$  3.5,  and we  repeat the  same
analysis for early-type  galaxies as we did above for  the case of C3.
The result  is presented in  Figure~\ref{fig:2PCCF_con}.  Although the
difference  on large  scales ($>1$Mpc)  becomes less  significant, the
stronger clustering  of barred galaxies still  persists on intermediate
scales, at around 500 kpc. As shown in \citet{Li-08a,Li-08b}, on these
scales the 2PCCF is dominated by  pairs of galaxies hosted by the same
dark  matter  halo.   Thus  the  difference in  2PCCF  between  barred
early-type galaxies and unbarred  early-type galaxies implies that the
two types of  galaxies are distributed in their  host dark matter halo
in different ways.  We will come back to this point in Section 4.4.

\subsection{Close neighbor counts}

A  problem with computing  the 2PCCF  on very  small scales  using our
spectroscopic sample  is the effect of fiber  collisions.  Although we
implemented  a  statistical  correction  that  is  proved  correct  on
average, it might still introduce systematic effects in our study.  In
this  subsection  we investigate  the  clustering  of  our samples  by
computing  the background-subtracted  neighbor  counts, $N_c(r_p<R_p)$,
which  is by  definition the  number  of galaxies  in the  photometric
reference sample  within the projected  radius $R_p$ of the  barred or
unbarred  galaxies,  with  the   effect  of  chance  projection  being
statistically corrected  (see Section 2).  When compared  to the 2PCCF
analyzed above, this quantity doesn't suffer from the fiber collision
effect on  small scales. It  also allows us  to explore the  effect of
fainter companion galaxies. For this  analysis we only consider our C3
samples where the dependence of clustering on mass, color and internal
structure of galaxies has been taken into account.

In Figures~\ref{fig:nc} and~\ref{fig:nc_ratio}, we plot $N_C$,
measured within a given value of
projected  radius $R_P$, for  barred and  unbarred galaxies in sample
C3. As in the previous  subsection, we make comparisons for barred and
unbarred galaxies  for given morphological type with their corresponding ratios of
$N_C$ for barred  galaxies relative to  those of unbarred
ones. Figure~\ref{fig:nc}, top to bottom, shows the results for the whole
sample, the early-type, and late-type subsamples, respectively, and in Figure~\ref{fig:nc_ratio},
we split present again the result for late-types as well as the result from
splitting the sample into strong and weak bars. Panels in both figures, from left to
right,  show  the  results  for different  apparent  $r$-band  limiting
magnitude applied  to the  photometric reference sample  ranging from
$r_{lim}=18$  for  the  leftmost   panels  to  $r_{lim}=21$  for  the
rightmost panels.  

In Figure~\ref{fig:nc} (top panels), where we compare the neighbor counts
around  all the  barred galaxies  and the  control sample  of unbarred
galaxies, we see  that barred galaxies have less  companions on scales of
$\la$50  kpc than unbarred galaxies,  while on scales  of $\ga$100 kpc,
barred   galaxies   present  more   neighbors   than  their   unbarred
counterparts.   When   we  split  the  galaxies   according  to  their
morphological  type, we  find that  the differences  on the  small and
large scales, found in the top panels for the barred galaxies as a whole,
are  essentially  contributed  by  late-type  and  early-type  galaxies
separately.  As  the second  row of panels  show, the  higher neighbor
count on large scales is  an effect seen only for early-type galaxies,
indicating  that barred,  early-type  galaxies are  located in  higher
density regions  than unbarred,  early-type galaxies of  similar mass,
color and  internal structure. This  confirms what we find  above from
the 2PCCF measurements, where the same sample of barred galaxies shows
higher clustering  amplitude on  intermediate scales than  the control
sample  of  unbarred galaxies.   When  the  comparison  is limited  to
late-type  galaxies,  the difference  on  large  scales  is no  longer
significant, while at  the smallest scales we see  clearly the drop in
$N_c$, suggesting  that the same effect  found for the  full sample is
driven by late-type galaxies.

Figure~\ref{fig:nc_ratio} presents the result for late-type galaxies only,
making the distinction between strong and weak bars.
The large  number of strongly barred  galaxies in our  sample helps to
improve  the  statistics and  reduce  the  statistical  errors in  our
measurements, but even with large error bars, if we look at the result
for weakly barred  galaxies we see that the  drop of average neighbor
counts is noticeable at even larger scales, up to a few hundred kiloparsecs. A
possible explanation for this effect  at larger scales for the case of
weakly barred galaxies might be  related with these systems being less
massive  than  those  galaxies  hosting  strong  bars  \citep{Lee-12b,
  Cervantes-Sodi-13},  which  would  make  them  less  stable  against
interactions with massive neighbors even at large separations.

Finally,  we notice that  in all  the cases  we have  considered, the
differences in  the neighbor count change  very little as  one goes to
fainter and fainter limiting  magnitudes. This implies that the excess
or  drop in  neighbor counts  seen above  is contributed  primarily by
relatively  bright  companions  with   $r<18$,  but  not  the  fainter
companions.

\subsection{Galaxy group catalog}

Our  results studying  the 2PCCF  and  average number  counts for  our
sample of  galaxies indicate that barred  early-type galaxies
present a higher clustering amplitude and a larger number of companion
galaxies on  scales from a  few hundred kiloparsecs  to several megaparsecs  when compared
with unbarred ones.  In particular, Figure~\ref{fig:2PCCF_con} reveals
a maximum at $\sim500$kpc in the ratio of the 2PCCF between barred and
unbarred  early-type  galaxies,  with  the ratio  decreasing  at  both
smaller and  larger scales.  This implies  that the 2PCCFs  of the two
samples are different not only in amplitude, but also in shape on these
scales. As shown  in \citet{Li-06b}, the amplitude of  2PCCF on scales
larger than  a few megaparsecs is  a direct measure  of the mass of  host dark
matter halos, while the shape of  the 2PCCF on a few $\times100$kpc is
sensitive to  the relative fraction of central  and satellite galaxies
in the  galaxy sample. Our  result here thus suggests  that early-type
barred galaxies are expected to  be found more frequently in satellite
galaxies than in central galaxies.

To further explore  this possibility we make use  of the group catalog
of \citet{Yang-07} in order to identify central and satellite galaxies
in our sample.   As described in the previous  section, after matching
our main sample  with the group catalog and  only keeping galaxies in
groups with three or more  members, our final sample reduces to nearly
2,000 galaxies.   From this reduced sample  we find that  the ratio of
central-to-satellite  systems   for  barred  and   unbarred  late-type
galaxies is very similar, about 27\% (see also \citet{Cervantes-Sodi-14}).
For early-type galaxies this
ratio is significantly different: 24\% for unbarred and only 16\% for
barred galaxies.  Although  the small sample size doesn't  allow us to
do a more detailed analysis, the  different central-to-satellite galaxy
number  ratio  for early-type  galaxies  is  already  telling us  that
early-type  barred galaxies  in our  sample are  indeed  more commonly
found as satellites in  galaxy groups, reinforcing our hypothesis from
the results using 2PCCFs and neighbor counts.

\subsection{Large-scale overdensity}

\begin{figure*}
  \begin{center}
    \epsfig{figure=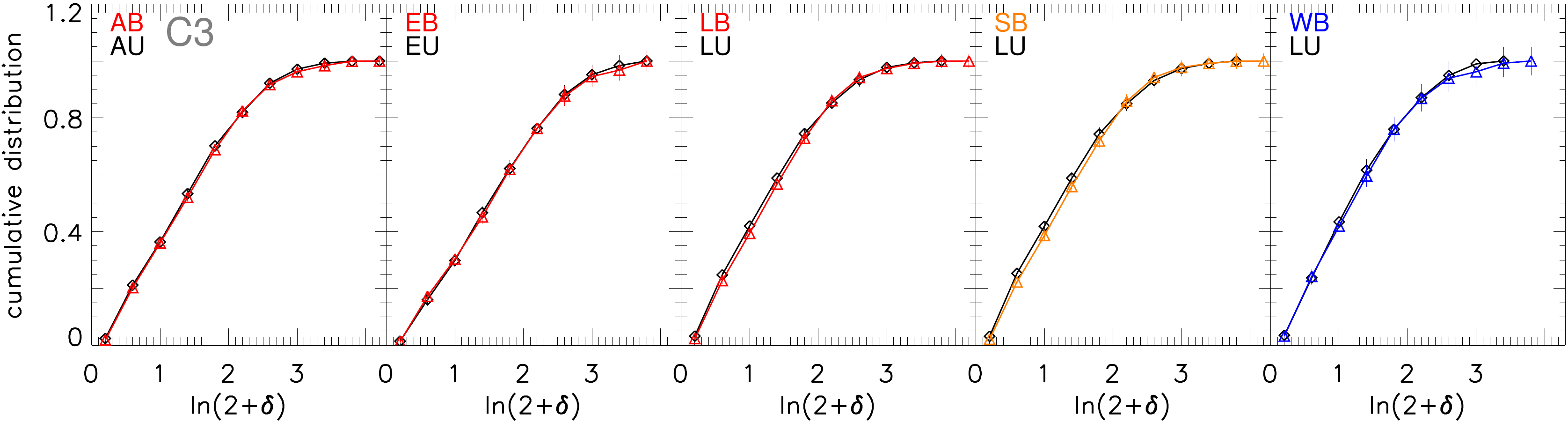,width=\textwidth}
  \end{center}
    \caption{Cumulative  distribution  of   barred  (color  line)  and
      unbarred (black line) galaxies  as a function of the overdensity
      parameter $\delta$ for the different subsamples of C3.}
  \label{fig:overdensity}
\end{figure*}

Finally,  we have  also examined  the overdensity  of  the large-scale
environment of our  barred and unbarred galaxies. For  this, we use the
overdensity  $\delta$  estimated  over   a  scale  of  $\sim3$Mpc  by
\citet{Jasche-10}  through a  reconstruction of  the three-dimensional
density  field   in  the  local   universe  based  on  the   SDSS  DR7
spectroscopic galaxy sample, as  described on Section 3.4. The results
are  shown  in  Figure~\ref{fig:overdensity},  where  we  compare  the
cumulative distributions  of the  different subsamples of  our control
sample  C3 as a  function of  ln(2 +  $\delta$). As  can be  seen, the
distributions of barred and  unbarred galaxies are almost identical in
all  cases. We  conclude that  the presence  of a  bar in  galaxies is
unlikely  correlated  with the  overdensity  of  local environment  on
scales of  $\sim3$Mpc.  This  is in good  agreement with what  we find
above from the 2PCCFs and close neighbor counts on large scales.

\section{Summary and discussion}

We have  presented an extensive  study of the environment  of galaxies
with  bars  in  the  low-redshift  universe,  which  makes  use  of  a
volume-limited  sample of $\sim$  30,000 galaxies  from the  SDSS with
visually determined    morphological     classifications    and    bar
identifications.  We use four  distinct statistics to characterize the
environment  of  the barred  galaxies  in  our  sample: the 
2PCCF $w_p(r_p)$  of our  galaxies with
respect to a spectroscopic reference sample, the background-subtracted
average  counts $N_c(R_p)$  of neighboring  galaxies in  a photometric
reference sample,  the membership  of our galaxies  in the  SDSS group
systems,  and the  overdensity $\delta$  of the  local  environment at
$\sim3$  Mpc  scale.   We  segregate  our  galaxies  into  early-  and
late-types to  study if differences  arise between the  two different
morphological   types.   For  the   late-type  subsample,   we  further
distinguish  two  types of  barred  galaxies, those with strong  and weak  bars,
according  to  their  size.   We  apply the  four  statistics  to  the
subsamples of  barred galaxies,  and compare  the results  to the
same statistics obtained for control samples of unbarred galaxies that
are  closely matched  to the  barred galaxy  samples in  stellar mass,
color, and internal structure in order to normalize out the dependence
of environment on these parameters.

 The first  thing we notice  measuring $w_p(r_p)$, once we  remove any
 dependence on  stellar mass, color and stellar  surface mass density,
 is  that the  clustering for  barred and  unbarred galaxies  are very
 similar, with only minor  differences. The differences arise when we
 split  the galaxies  in our  sample according  to  their morphological
 types.  For the case of late-type galaxies, the 2PCCF does not depend
 on the presence  or absence of a bar,  confirming the previous result
 by  \citet{Li-09}.  The  barred early-type  subsample shows  a higher
 amplitude for  $w_p(r)$ on scales from a  few hundred kiloparsecs to  1 Mpc, when
 compared with unbarred galaxies,  very similar to the result reported
 by \citet{Skibba-12},  who found a  higher likelihood for  barred and
 bulge-dominated galaxies on denser  environments on scales of 150 kpc
 to  a   few  megaparsecs,  than  unbarred,  disk-dominated   ones.  At  these
 intermediate  scales, the  correlation function  is dominated  by the
 one-halo term,  which would indicate that  barred early-type galaxies
 are more frequently found as  satellite systems. To explore this idea
 further, we  made use of the \citet{Yang-07} group  catalog to identify
 our galaxies  in groups as  centrals or satellites, finding  that the
 percentage of barred early-type galaxies was significantly higher for
 satellites compare with the percentage for centrals. In contrast, the
 percentage of  barred late-type galaxies  is the same for  central as
 for satellite galaxies, in agreement with recent results by \citet{Cervantes-Sodi-14}.
 
 With  our barred  early-type  sample composed  by  S0 galaxies,  this
 result is also in agreement with \citet{Barway-Wadadekar-Kembhavi-11}
 who found that S0 galaxies  in clusters present a higher bar fraction
 than      their     counterparts      in      the     field,      and
 \citet{Lansbury-Lucey-Smith-14}, who  analyzed a reduced  sample of S0
 galaxies in  Coma and found  an increase in  the bar fraction  from the
 outskirts of  the cluster toward  the core. A possible  framework to
 explain this trend is that  these S0 galaxies were originally spirals
 that were transformed by  ram pressure stripping with the intracluster
 medium,  loosing  their cold  interstellar  gas  to the  environment,
 causing a fading of the disk and subsequently a morphology change, but
 preserving their stellar features such as the bar. This mechanism is
 expected    to   work    only   for    faint    lenticular   galaxies
 \citep{Boselli-Gavazzi-06,Barway-09},  which are  not  able to  retain
 their  gas,  while  bright  S0s  presumably  form  through  other
 processes (e.g., mergers).  Interestingly, our barred lenticulars are
 mostly  faint  galaxies, with  M$_r  >  -21.5$.   With a  higher  bar
 fraction  for spirals  than  for lenticulars  \citep{Laurikainen-09},
 this  could explain  the higher  amplitude found  for  the correlation
 function in our  sample of barred early-type galaxies  on scales that
 correspond to satellite systems of more massive halos.
 
 Given that  our estimate  for the 2PCCF  suffers the effect  of fiber
 collisions at  small scales, we  turned to close neighbor  counts to
 study the  clustering of  barred galaxies at  small scales.   In this
 case,  it is  the barred  late-type subsample   that  shows a
 systematic trend of  having an average lower number  of neighbors on
 scales of $\le$50 kpc when  compared with the control sample of unbarred
 late-type galaxies.  The  effect seems more dramatic for  the case of
 weakly  barred  galaxies,  where  the deficiency  appears  at  larger
 scales  close to  100 kpc.   This  finding may  indicate that  tidal
 interactions  destroy   and/or  disfavor  bar   growth  instead  of
 enhancing it  as expected from  results of numerical  simulations. As
 weak bars are  more commonly found in fainter,  less-massive galaxies
 than strong bars  \citep{Lee-12b,Cervantes-Sodi-13}, the fact that we
 find the effect of nearby companions at larger scales for the case of
 weak  bars may  be  not unexpected:  due  to their  lower mass  their
 gravitational potential is  also weaker and they are  more fragile to
 disturbances than their more  massive counterparts.  In addition, the
 decline  of  the  average  close  neighbor count  around  the  barred
 galaxies  in  our sample  is  found to  be  weakly  dependent on  the
 limiting  magnitude  of the  photometric  sample,  implying that  the
 effect is  primarily driven by relatively  bright companion galaxies,
 with little contribution from companions fainter than $r=18$.
 
 Our result echoes previous studies  that report a decrease on the bar
 fraction  for  disk  galaxies  when  the separation  to  the  nearest
 neighbor becomes  smaller than  0.1 times the  virial radius  of the
 neighbor  \citep{Lee-12a}, a  lower fraction  of barred  galaxies in
 close     pairs    when     compared    with     isolated    galaxies
 \citep{Mendez-Hernandez-11}  and  a  decrease  on the  likelihood  of
 identifying    a    bar    in    pairs   with    small    separations
 \citep{Casteels-13}.   In  any  case,  we only  find  indications  of
 suppressed formation/growth of stellar bars in galaxies that might be
 undergoing some kind  of interaction with a close  neighbor, with no
 evidence  of environmental  stimulation, in  opposition  with earlier
 studies \citep{Elmegreen-Elmegreen-Bellin-90}.
 
 Finally, we do not find  any dependence of the likelihood of galaxies
 hosting  bars on  the large-scale  environment, as  accounted  by the
 overdensity parameter $\delta$, estimated through a reconstruction of
 the three-dimensional  cosmic web posterior \citep{Jasche-10}.  This result suggests
 that  galactic  bars are  not  obviously  linked  to the  large-scale
 structure of the universe.

\acknowledgments
The authors  thank the anonymous referee for comments that improved the paper.
We are grateful to Changbom  Park for providing the galaxy sample used
in this study and Ramin Skibba, Edmond Cheung,  and Jairo Mendez-Abreu
for helpful discussions. 
This work is sponsored by NSFC (grant Nos. 11173045, 11233005, 11325314,
11320101002) and the Strategic Priority Research Program ``The Emergence of
Cosmological Structures'' of CAS (grant No. XDB09000000).
Funding for  the SDSS and SDSS-II  has been provided by  the Alfred P.
Sloan Foundation, the Participating Institutions, the National Science
Foundation, the  U.S.  Department of Energy,  the National Aeronautics
and Space Administration, the  Japanese Monbukagakusho, the Max Planck
Society,  and the Higher  Education Funding  Council for  England. The
SDSS Web  site is  http://www.sdss.org/.  The SDSS  is managed  by the
Astrophysical    Research    Consortium    for    the    Participating
Institutions. The  Participating Institutions are  the American Museum
of  Natural History,  Astrophysical Institute  Potsdam,  University of
Basel,  University  of  Cambridge,  Case Western  Reserve  University,
University of Chicago, Drexel  University, Fermilab, the Institute for
Advanced   Study,  the  Japan   Participation  Group,   Johns  Hopkins
University, the  Joint Institute  for Nuclear Astrophysics,  the Kavli
Institute  for   Particle  Astrophysics  and   Cosmology,  the  Korean
Scientist Group, the Chinese  Academy of Sciences (LAMOST), Los Alamos
National  Laboratory, the  Max-Planck-Institute for  Astronomy (MPIA),
the  Max-Planck-Institute  for Astrophysics  (MPA),  New Mexico  State
University,   Ohio  State   University,   University  of   Pittsburgh,
University  of  Portsmouth, Princeton  University,  the United  States
Naval Observatory, and the University of Washington.

\label{lastpage}
\end{document}